# Hybrid Classical-Quantum Solutions to Accelerate the Adoption of Quantum Computing

Advait Deshpande
The Open University
United Kingdom

Luciano Baresi
Politecnico di Milano
Italy

Arosha K. Bandara
The Open University
United Kingdom

Despite the promise and potential of quantum computing, most of the current day developments of quantum computing have several limitations for relevant practical use. As a result, hybrid classical-quantum computing has emerged as a viable solution. This paper discusses the software aspects of hybrid classical-quantum computing solutions to understand and examine how hybrid computing solutions can be expected to work in practice.

This paper examines existing software solutions (cloud applications, middleware frameworks, and API gateways) that can enable classical and quantum computing to work together. For practitioners the paper provides a curated list of algorithms, full stack libraries, simulators, and cloud service providers that are likely to be central to the development of any hybrid computing solutions. We conclude by identifying an agenda for future research on classical-quantum hybrid software engineering, and some near-term priorities for development of functional, robust hybrid software systems. The paper argues that due to the deep dependency of quantum execution on classical pre- and post-processing, hybrid computing may represent a near-permanent outcome of classical-quantum computing rather than a transitional workaround.

***Keywords:*** *Hybrid classical-quantum computing, Hybrid computing software stacks, Quantum software engineering, NISQ, Functional and robust hybrid software systems*

## 1 INTRODUCTION

Quantum computing refers to a type or classification of computing based on quantum mechanics where information is encoded and processed by controlling quantum states [22]. In contrast to classical digital computing, quantum computing can perform calculations and evaluate extensive combinations of hypotheses simultaneously instead of sequentially [33]. Consequently, quantum computers have the potential to solve problems in far fewer steps (i.e. faster) than classical computers, also called *Quantum speedup*, and could revolutionise computing and by extension many aspects of human life.

The breadth of potential applications illustrated in Table 1 highlights quantum computing's potential reach and the nature of advantages it offers. Quantum speedup is not a uniform property that applies equally to all problem types. It depends critically on matching the right algorithmic approach to the underlying mathematical structure of the problem. Combinatorial optimisation problems, such as those found in manufacturing scheduling and supply chain logistics, are well-suited to the Quantum Approximate Optimisation Algorithm (QAOA), which exploits quantum superposition to explore solution spaces more efficiently than classical search heuristics [15]. Simulation problems, such as molecular modelling in chemical engineering and pharmaceutical R&D, align naturally with the Variational Quantum Eigensolver (VQE), which leverages entanglement to represent quantum-mechanical states that are computationally intractable for classical hardware [9]. Cryptanalytic problems, by contrast, derive their quantum advantage from Shor's algorithm, which exploits quantum Fourier transforms to achieve exponential speedup for integer factorisation [36]. This algorithmic diversity has a direct structural consequence for the software stack: because different problem domains require fundamentally different quantum algorithmic approaches, the software infrastructure for hybrid computing cannot be a single unified framework. Instead, it is likely to remain a collection of problem-specific workflows, each with distinct requirements for classical pre-processing, quantum circuit construction, and post-processing of probabilistic outputs. The architecture of the hybrid software stack therefore mirrors the diversity of quantum advantage types shown in Table 1.

As described in Table 1, the quantum advantage types can be understood as:

- Optimisation: quantum speedup for combinatorial and continuous optimisation problems via algorithms such as QAOA and VQE [12, 15].
- Simulation: exponential speedup for simulating quantum-mechanical systems (molecules, materials) via quantum phase estimation and VQE [9, 28].
- Sampling: quadratic-to-exponential speedup for drawing samples from probability distributions, e.g. in risk modelling and Monte Carlo finance [26, 35].
- Cryptanalytic speedup: Shor's algorithm provides exponential speedup for integer factorisation and discrete logarithms, breaking Rivest–Shamir–Adleman (RSA)/ Elliptic-Curve Cryptography (ECC)-class cryptography on fault-tolerant hardware [36]. Post-quantum cryptography concerns resistance to this attack, rather than a quantum advantage.

Table 1: Potential applications of quantum computing and expected quantum advantage type [3] [6]

| Sector / industry / area | Potential applications of quantum computing | Quantum advantage type |
|---|---|---|
| Advanced manufacturing | • Combinatorics problem-solving to infer causal links in process failures<br>• Predictive modelling based on unbalanced data on profit margin and return on assets (ROA) | Optimisation |
| Aerospace design and engineering | • Surrogate modelling of partial differential equations<br>• Wingbox design optimisation | Simulation; Optimisation |
| Automotive | • Design optimisations for electric drives using numerical simulation<br>• Performance guarantees for eVehicle batteries<br>• Robot path optimisation for production robots<br>• Optimisation of labour shift assignments | Optimisation; Simulation |
| Banking / Finance | • Portfolio optimisation, foreign exchange arbitrage, and credit scoring<br>• Derivatives pricing and risk management<br>• Simulation of market predictions | Optimisation; Sampling |
| Chemical engineering (incl. pharmaceuticals) | • Material discovery and new drug development<br>• Development of new proteins<br>• Target identification, drug design and toxicity testing<br>• Chemical simulations as part of pharmaceutical R&D | Simulation |
| Cybersecurity | • Quantum Key Distribution (QKD) for algorithmic encryption<br>• On-chip quantum encryption hardware<br>• Post-quantum cryptography | Cryptanalytic speedup |
| Material science | • Modelling polymers, solids, molecules at high precision | Simulation |

| Sector / industry / area | Potential applications of quantum computing | Quantum advantage type |
| --- | --- | --- |
| | • Prediction of chemical reactivity in Molecular Quantum Chemistry<br>• Simulation of the dynamics of molecules<br>• Chemistry calculation for battery research | |
| Healthcare | • Diagnostic assistance to patient care<br>• Personalised interventions and medical treatments at scale<br>• Optimisation of insurance premiums and pricing | Optimisation; Sampling |
| Supply chain | • Demand capacity matching and production planning<br>• Optimisation of supply chain processes<br>• Transportation cover and logistics of various operations<br>• Supply chain lot sizing | Optimisation |
| Transport | • Optimisation of vehicle utilisation in a transport network | Optimisation |

Despite this promise and potential of quantum computing however, most of the current day developments of quantum computing have several limitations when their use in practice is considered [30]. Quantum bits i.e. qubits which hold the basic piece of information in quantum computers, are highly error-prone and cannot retain their qubit state in a persistent manner. In other words, quantum computers lose input data, may process incorrect data, and therefore may produce more errors than useful information. Qubits are also highly unstable and lose information when they interact with the external environment and therefore require cryogenic (i.e. sub-zero) temperatures to be operational. The cost of operating functional quantum computers is therefore prohibitively expensive. As of 2025–26, most state-of-the-art quantum platforms can consistently achieve error rates of $10^{-3}$ (0.001%) for physical qubits, and need to achieve error rates of $10^{-15}$ (0.000000000000001%) for physical qubits before they can match the capabilities of modern-day classical digital computers [17]. Some experts suggest that such improvements are years (if not decades) away before quantum computing can be meaningfully useful. In response, the prospect of hybrid quantum-classical computing has emerged as a viable solution [30].

Hybrid classical-quantum computing (hereafter hybrid computing), aims to harness the perceived capability of quantum computers in solving highly complex problems in conjunction with the maturity, flexibility, and accessibility of classical computers. This is analogous to modern classical computers in which the Graphics Processing Unit (GPU) is optimised to perform vector operations, used for graphics rendering and other computational tasks like machine learning while the Central Processing Unit (CPU) is utilised for performing a wider range of tasks quickly but with limited concurrency. In the hybrid computing paradigm therefore quantum computers can be designated to execute a small, specialised part of the workflow with the classical computers performing the larger total workflow [30] (See Figures 1 and 2 for reference). As the key performance issues of error correction, decoherence, and noise reduction are addressed, and the quantum computing technology becomes more robust, the role of classical computing technologies can be expected to reduce. However, current day quantum computers have limited native support for imperative programming constructs like branching and iteration due to their use of the circuit model which is more like functional programming rather than imperative programming. In hybrid computing systems, classical computers enable imperative programming approaches through real-time feedback and mid-circuit measurements during quantum execution, and manage quantum operations using classical control systems [23]. As a result, classical computers are expected to remain at the core of the hybrid computing approach in the near-term.

Given the high cost of developing, producing, and maintaining quantum computers, Quantum-as-a-Service models are expected to enable practical hybrid computing solutions. By delivering quantum computing through cloud-based platforms such as Azure Quantum, Braket, and IBM Quantum, big tech companies such as Microsoft, Amazon, and IBM are offering early adopters of quantum computing the opportunity to develop and deploy hybrid quantum-classical software solutions. Whether such solutions can scale to achieve the commercial success, modularity, maintainability, and reusability of current day service-oriented computing is however unclear as of 2025–26 [26].

In this context, this paper makes three contributions to the existing research and analyses of hybrid quantum-classical computing.

- First, it examines the software stack of hybrid systems as a subject in its own right, rather than treating software infrastructure as an adjunct to quantum algorithm research. The predominant focus of the quantum computing literature is on algorithmic correctness and hardware performance; the engineering challenges of the software layer that connects classical and quantum components (i.e. orchestration, verification, interoperability, and workflow management) have received comparatively little systematic attention.
- Second, the paper maps existing solutions against the specific limitations they address, producing a structured gap analysis rather than a descriptive survey of available tools. This mapping, presented in Table 3, reveals which limitations have at least partial solutions in the current landscape and which remain structurally unaddressed.
- Third, the paper argues that hybrid computing is not a transitional state to be superseded by fault-tolerant quantum computing, but a structural feature of classical-quantum computing for the foreseeable future. This argument has implications for how the field should prioritise research investment not only in hardware capability, but also in the software engineering foundations that will determine whether hybrid systems can be deployed reliably at scale.

## 2 HYBRID CLASSICAL-QUANTUM COMPUTING/SYSTEMS

There are different approaches to hybrid computing and how the workflows of classical and quantum computers can be utilised to optimise solutions to certain problems. In consequence, there are different schools of thought about how hybrid computing capabilities should be assembled, and optimised for problem-solving. Phillipson et al [30] distinguish hybrid computing in two main (but not mutually exclusive) categories:

- **Vertical hybrid computing** in which all activities required to control and operate a quantum circuit take place on a quantum computer, and the classical computer similarly does compilation of its own workflows. This approach is primarily about the integration of the quantum and classical hardware and is typically application-agnostic.
- **Horizontal hybrid computing** in which all operational activities require the use of quantum and classical computers in an interdependent manner. This approach is application-dependent and primarily about the algorithms.

The distinction between vertical and horizontal hybrid computing is not merely taxonomic. It has practical consequences for where software stack problems will be most acute. Vertical hybrid computing is, in essence, what all current quantum cloud services implement: a classical computer submits a compiled circuit to a remote quantum processor, execution takes place on the Quantum Processing Unit (QPU), and results are returned to the classical layer for interpretation. This model requires relatively modest software integration; the classical and quantum components interact at a well-defined boundary, and the classical orchestration burden is limited to circuit compilation and result parsing. Horizontal hybrid computing is a fundamentally different proposition. Variational quantum algorithms (VQAs) such as VQE and QAOA require genuine interleaving of classical and quantum computation: a classical optimiser iteratively adjusts the parameters of a parameterised quantum circuit based on measurement outcomes, repeating this loop until convergence. This pattern demands low-latency, tightly coupled communication between the classical and quantum layers, real-time feedback during quantum execution, and co-scheduling of classical and quantum compute resources. It is horizontal hybrid computing that places the most stringent demands on the software stack, and where the current tooling has scope to mature.

Figures 1 and 2 below depict the vertical and horizontal hybrid computing approaches at a very high-level. Solid line borders depict classical components and dashed line borders depict quantum components.

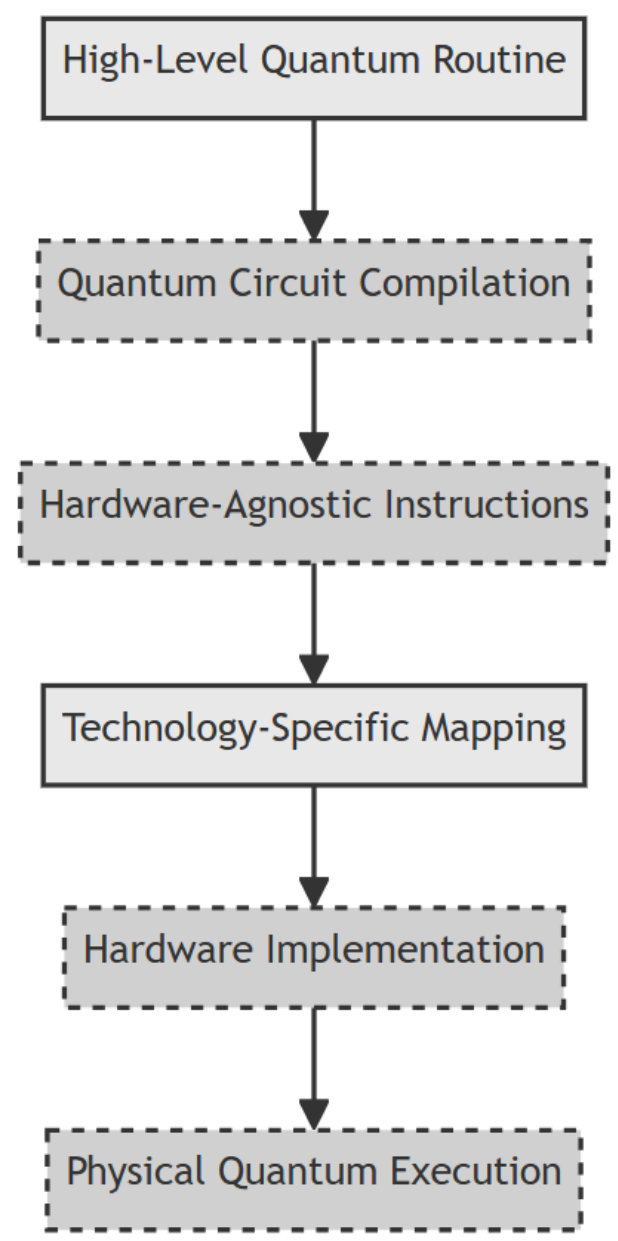

Figure 1: Vertical hybrid computing (based on [30])

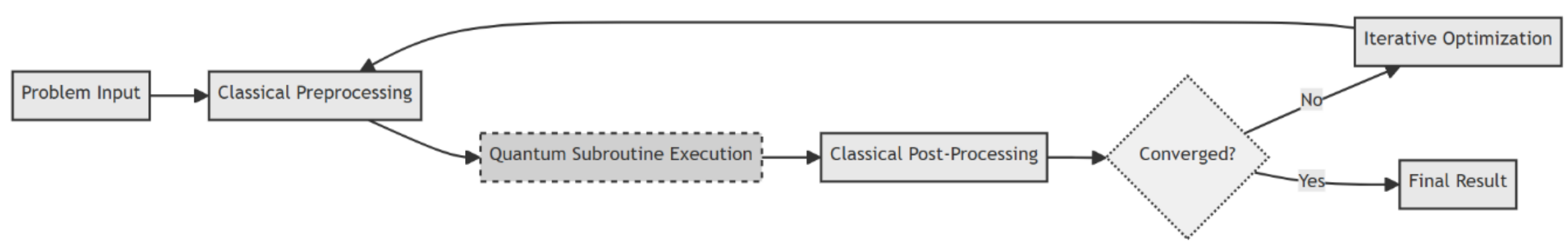

Figure 2: Horizontal hybrid computing (based on [30])

This discussion focuses on the software stack of hybrid computing solutions to understand and examine how hybrid computing solutions can be expected to work in practice.

Endo [13] and Endo et al [14] contend that since classical computers will execute a large part of the workflows in the first iterations of hybrid quantum solutions, there is likely to be less emphasis on the issues of decoherence and noise reduction in the performance of classical computers. From this perspective, Variational Quantum Algorithms (VQA) have emerged as an important approach to hybrid computing. In VQAs, the classical computer is used to train a parameterised quantum circuit, which provides a framework for solving a wide array of tasks such as solving linear systems of equations and machine learning [9]. The Variational Quantum Eigensolver (VQE) algorithm, considered the very first VQA, is seen as an example of hybrid computing in a noisy environment (i.e. effectively the classical computer performs most of the workflows).

From a practical perspective, this results in two types of workflows to ensure effective division of computational tasks between quantum and classical computing resources:

- **Algorithms with pre or post-processing of a quantum computation on a classical computer.** With these algorithms, classical computers either pre-process the problem into a task (or set of tasks) which the quantum computer can recognise, and/or post-process the data returned by quantum computer so it can be effectively used to provide a correct answer [8] [1]. Examples of this approach are Shor's algorithm [36] and Simon's algorithm [37] that use classical post-processing.
- **Algorithms that perform multiple iterations of quantum and classical computations,** whereby the output of the quantum computation is improved in each iteration until the result reaches the required accuracy (i.e. convergence as shown in Figure 2). An example of this approach is the quantum approximate optimisation algorithm (QAOA), which is a kind of VQA [15].

Since most current day quantum computers have their quantum mechanical systems controlled by some form of classical computing workflows, Calude et al [8] and Allen and Calude [1] contend these are hybrid computers i.e. a full-stack vertically integrated quantum computer functioning without the intervention of a classical computer does not yet exist. Given the dependency of quantum computers on pre-processing of data through classical computers to prepare it for quantum algorithms or post-processing of the data from quantum algorithms for further analysis, Edwards [12] suggests that quantum application may always remain hybrid. This view is supported by Callison and Chancellor [7], who demonstrate that hybrid classical-quantum algorithms remain the dominant paradigm not only in the current Noisy Intermediate-Scale Quantum (NISQ) era but in the foreseeable future beyond it.

Fault-tolerant quantum computing, if achieved, could eventually eliminate the dependency on classical co-processing by enabling quantum hardware to execute deep, error-corrected circuits autonomously. However, there are three structural reasons due to which it is unlikely to permanently displace the classical computing paradigm entirely.

- First, classical computers retain decisive advantages in data input/output, random-access memory, and sequential processing throughput that quantum hardware is not architecturally designed to match. The von Neumann model excels at the data-management tasks that quantum algorithms require at their periphery: loading problem instances, storing intermediate results, and post-processing probabilistic outputs.
- Second, economic and deployment constraints mean that quantum hardware will remain a shared, remotely accessed resource for the foreseeable future. Cloud-based Quantum-as-a-Service models are not merely a stopgap; they reflect the practical reality that quantum processors are expensive, cryogenically maintained, and institutionally concentrated. Accessing them through classical interfaces is therefore an architectural given, not a temporary inconvenience.
- Third, the translation of real-world problem data into quantum-compatible representations is intrinsically the domain of classical computing. Examples include encoding classical data into qubit states, mapping optimisation objectives to Quadratic Unconstrained Binary Optimisation (QUBO) formulations, and interpreting measurement distributions as actionable outputs. Fault-tolerant hardware does not make these translation tasks quantum-executable. However, it enables longer coherent computation once the data has been encoded. Callison and Chancellor [7] support this position to contend that hybrid algorithms will retain their structural advantages beyond the NISQ era.

Effectively, for optimal outcomes the resources of classical and quantum computers need to be allocated to their strengths. For example, classical computing is deployed for data processing and analysis, and quantum computing is employed for optimisation and simulation problems (see Table 1). Given that quantum computers require extensive physical resources for error correction at this stage of their development, VQAs can be a possible practical solution since they can enable certain tasks to be reliably executed with current quantum technology.

Table 2 provides a curated, partial list of algorithms, full stack libraries, simulators, and cloud service providers which are likely to be central to the development of any hybrid computing solutions. This list is partially based on resources provided by the Quantum Open Source Foundation (QOSF) [32], curated based on availability of documentation for practitioners (including tutorials, ease-of-learning, and open source nature), long-term funding prospects, possible integration of hardware and software workflows, and suitability for hybrid solution development.

As used in Table 2, the maturity labels should be understood as follows:

- Production-ready = stable public releases with Service Level Agreements (SLAs), active maintenance, and documented enterprise use;
- Early-stage = active development, Application Programming Interface (API) instability, primarily research/prototype use;
- Research = proof-of-concept or unmaintained.

Table 2: Curated list of hybrid computing resources with maturity, language integration, and hybrid workflow support [32]

| Type | Examples | Maturity | Classical language integration | Native Hybrid Workflow Support |
| --- | --- | --- | --- | --- |
| Quantum algorithms | • Arline Quantum<br>• Grove / QuantumFlow (Rigetti)<br>• PennyLane (Xanadu)<br>• Qiskit Nature, Optimization, ML, Finance (IBM)<br>• Quantum Katas (Microsoft) | Early-stage to Production-ready | Python-native | Partial[1] — algorithmic primitives designed for hybrid VQA/QAOA loops; orchestration requires external framework |

[1] PennyLane is rated Production-ready for ML/VQA workflows [31]; Arline Quantum and Grove are Early-stage research tools. Qiskit reached stable 1.0 release in February 2024 [27].

| Type | Examples | Maturity | Classical language integration | Native Hybrid Workflow Support |
|---|---|---|---|---|
| Full-stack libraries | • Braket (Amazon) | Production-ready | Python-native; REST API | Yes — native hybrid job execution via Amazon Braket Hybrid Jobs[2] |
| | • Cirq (Google) | Early-stage | Python-native | Partial[3] — supports parameterised circuits and VQA patterns; no built-in classical orchestration layer |
| | • Forest / pyQuil (Rigetti) | Early-stage | Python-native | Partial[4] — supports QVM/QPU execution loops; requires manual classical integration |
| | • Ocean (D-Wave) | Production-ready | Python-native | Yes — hybrid solver service natively integrates classical and quantum annealing |
| | • Q# (Microsoft) | Early-stage | Multi-language (Python, .NET, C#) | Partial[5] — hybrid via Azure Quantum; no standalone classical orchestration |
| | • Qiskit (IBM) | Production-ready | Python-native; REST API | Yes — Qiskit Runtime Primitives natively execute |

[2] Amazon Braket Hybrid Jobs provisions classical compute alongside QPU access, enabling managed hybrid execution [39].
[3] Partial (Cirq, Forest, Q#): supports quantum circuit execution and parameterised loops but requires the developer to implement the classical optimisation loop manually [25] [31].
[4] ibid
[5] ibid

| Type | Examples | Maturity | Classical language integration | Native Hybrid Workflow Support |
|---|---|---|---|---|
| | | | | parameterised hybrid circuits on IBM hardware[6] |
| Quantum simulators | • Intel Quantum Simulator<br>• Liqui\|> (Microsoft)<br>• QVM (Rigetti)<br>• Qiskit Aer (IBM)<br>• QuEST (Oxford) | Research to Production-ready | Python-native (most); C/C++ (QuEST, Intel) | Partial[7] — simulators emulate quantum execution but classical orchestration must be implemented externally |
| Cloud service providers | • IBM Q Experience (IBM Quantum) | Production-ready | Python-native; REST API | Yes — Qiskit Runtime delivers native hybrid primitives (Sampler, Estimator) |
| | • Google Quantum AI | Early-stage | Python-native | Partial[8] — hybrid workflows via Cirq; no fully managed hybrid execution service |
| | • Xanadu Quantum Cloud | Early-stage | Python-native | Yes — PennyLane's device-agnostic autodiff natively supports hybrid training loops |
| | • Forest / QCS (Rigetti) | Early-stage | Python-native | Partial — QPU + QVM access; manual classical loop integration required |
| | • Amazon Braket | Production-ready | Python-native; REST API | Yes — Hybrid Jobs service with managed classical |

[6] Qiskit Runtime Primitives (Sampler, Estimator) abstract hybrid execution on IBM hardware natively [20] [42].

[7] Maturity varies: Qiskit Aer is Production-ready; QuEST and Intel QS are research-grade C/C++ simulators; Liqui|> is no longer actively maintained.

[8] See footnote 3

| Type | Examples | Maturity | Classical language integration | Native Hybrid Workflow Support |
|---|---|---|---|---|
| | | | | compute alongside QPU access |
| | • Azure Quantum (Microsoft) | Production-ready | Multi-language (Python, Q#, .NET) | Yes — hybrid workflows via Azure Quantum resource estimator and Q# integration |
| | • D-Wave | Production-ready | Python-native | Yes — Leap hybrid solver service is purpose-built for hybrid classical-quantum annealing |

Although the perceived superiority of quantum computing for optimisation and simulation problems may be an enabler for adoption of hybrid computing, as the above list shows, the extent to which available commercial quantum solutions (most of which are cloud-based) offer sufficient performance gains over scaled-up classical computing hardware is unclear at this stage.

## 3 HYBRID COMPUTING SOLUTIONS IN PRACTICE

Since all current quantum systems are hybrid in practice, this section considers how classical and quantum computing can work together, instead of having the classical parts subordinated to the quantum computations. The goal is to understand how these different computations can cooperate instead of having one that works for the other. To this end, there have been different proposals, but the concrete examples have been very limited in terms of addressed problems and the size of the resulting systems.

One of the first solutions that tried to move in this direction was TOSCA4QC [41], an extension to the Topology and Orchestration for Cloud Applications (TOSCA [5]), which is a general-purpose standard for the management and deployment of cloud-based applications, that tries to cope with the specialities introduced by Quantum computing. The work stays at a high-level, and the extensions aim to support the two stacks needed to run both software types independently. The example application is very simple and comprises a toy web app that interacts with a Quantum application, implemented as a Qiskit app, that offers Shor's algorithm for factorising numbers in polynomial time.

The same group also introduced Quokka [4], a service ecosystem that provides dedicated services for the implementation of the different tasks of variational quantum algorithms as workflows. While the emphasis is on variational algorithms, this work can be seen as a first step towards the workflow-based composition of both classical and quantum components. Quantum Function-as-a-Service (QFAAS) [28] seems to move even a step further and claims it offers a 'full-stack and unified quantum serverless platform by integrating multiple well-known quantum software development kits (Qiskit, Q#, Cirq, and Braket), quantum simulators, and cloud providers (IBM Quantum and Amazon Braket)'. The idea of quantum function-as-a-service (FaaS) seems the pre-condition for being able to exploit the FaaS paradigm as an aggregator of heterogeneous computation models. The framework mainly allows one to wrap calls to quantum APIs through cloud functions, but there is nothing about integrating classical and quantum functions to make them cooperate, and the only two provided examples refer once again to Shor's algorithm and random number generation. Despite their promise, many anticipated quantum software engineering features are not yet needed in practice.

There have also been proposals of middleware frameworks [16] for integrating quantum and classical computing into a heterogeneous computation architecture with the claimed goal of enhancing the overall performance and efficiency in solving a wide range of problems. The framework is implemented on top of Ray.io [2], but again the proposed case studies refer to circuit knitting and Quantum Machine Learning (QML).

A similar role is played by the Quantum API Gateway [18] whose aim is to allow users to call classical and quantum components (services) in the same way. The Quantum API Gateway is a prototype that could be used to exploit the services provided by the Amazon Braket platform. However, the goal of their evaluation was not the integration between heterogeneous components, but the quantification of the delay introduced by the gateway when interacting with a Quantum component.

Although there have been other attempts to exploit well-known principles borrowed from distributed systems and software engineering to pave the ground to the integration, examples of applications that exploit it proficiently in practice remain limited.

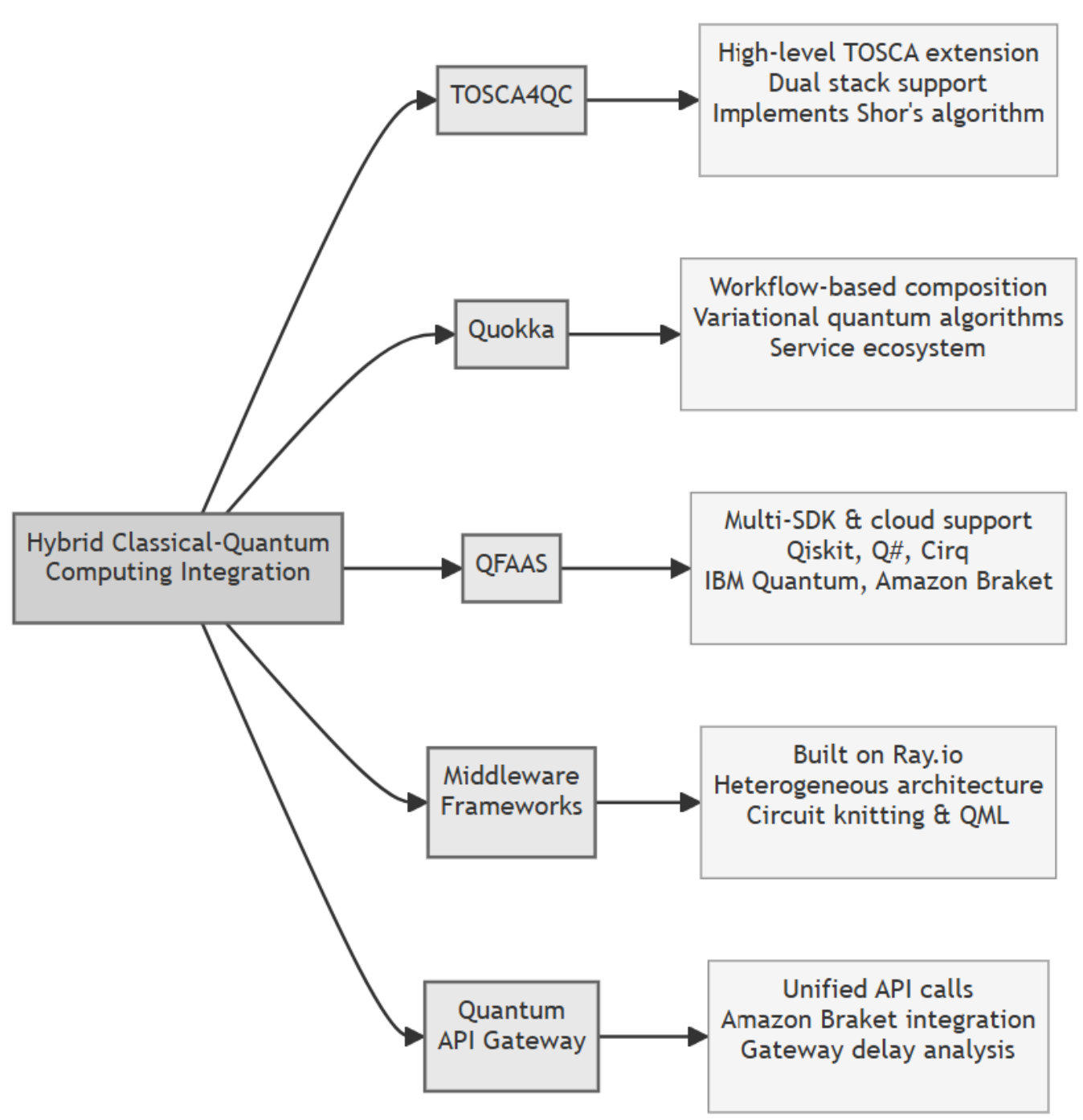

Figure 3: Overview of existing hybrid classical-quantum integration solutions

## 4 CURRENT LIMITATIONS

The main limitation of quantum computing which hybrid computing is supposed to address is that in the near-term quantum computing offers insufficient performance gains compared to classical computing for any industrial / business-relevant, mission critical applications. Due to the extreme sensitivity of qubits to their environment, qubit decoherence which causes the qubits to lose their quantum properties is a critical, path dependent challenge which quantum computing needs to overcome on the hardware stack. Effectively, current day quantum computers produce far more errors than accurate results. As the errors in quantum computing accumulate, the quality of computational outputs degrades making quantum computers impractical for even niche use cases. This suggests the prospect of universal everyday use cases is unlikely without significant breakthroughs in hardware and software capabilities [10].

On the software stack, quantum algorithms are far more complex than classical algorithms and therefore require developers to approach computational problems in original ways which do not fit the classical Von-Neumann computational approaches. This highlights the need for new programming languages, compilers, and optimisation tools to make effective use of quantum computing capabilities. In particular, such tools need to be capable of efficient and reliable transfer of data between classical and quantum computing stacks given that a universal quantum computer capable of replacing classical computers will need to overcome the limitations of current day materials and computational techniques.

By design, hybrid computing is not intended to address all these challenges facing quantum computing. The key challenges to hybrid computing are about ensuring that the hardware and software workflows deliver accurate outputs in a consistent, replicable manner. However, the invocation and execution of quantum-as-a-service models is currently closely aligned with the quantum processors [35]. There is significant variation in the way current day quantum computers provide outputs of the executed process. These outputs are often likely to be error-prone depending on the underlying quantum hardware and qubit topology, contributing to strong coupling between the service and underlying hardware which does not reliably provide intermediate verification of outputs. Given the lack of interoperability between cloud platforms, there is significant technical overhead to performance verification of software, benchmarking of hardware, and validation of hardware behaviour using numerical simulation [24].

Table 3: Summary of key limitations and mapping to existing solutions or identified gaps

| | Type of limitation | Description of limitation | Existing solution/gap |
|---|---|---|---|
| **Quantum computing** | Hardware | Insufficient performance gains compared to classical computing for industrial/mission-critical applications | VQA / QAOA approaches (partially) defer performance requirements to near-term hardware [9] [8] [15] [7] |
| | | Qubit decoherence causes the qubits to lose their quantum properties | Error mitigation via classical post-processing in hybrid loops [14] [9] |
| | | Produce far more errors than accurate results | No current solution: Decoherence remains the fundamental hardware bottleneck |
| | | Quality of computational outputs degrades over time | No current solution: Fault-tolerant quantum computing is under development [17] |
| | Software | Quantum algorithms far more complex than classical algorithms | Domain-specific languages (Liquid, Q#, Qiskit) have begun to address this [39] [38] [20] [42] |
| | | Developers must abandon classical Von-Neumann computational approaches | No current solution: New computational paradigm requires deep re-training |
| | | New programming languages, compilers, and optimisation tools required | Partial solution: Qiskit, PennyLane, and Cirq provide algorithm libraries; standardisation efforts ongoing [32] [31] |
| **Hybrid quantum-classical computing** | Hardware | Quantum-as-a-service tightly coupled to underlying quantum processors; bespoke per-platform | Partial solution: TOSCA4QC [41] and Quokka [4] address orchestration; portability gaps remain |
| | | Strong coupling between service and hardware; no intermediate verification of outputs | No current solution: Verification is an open research problem [24] |
| | | Lack of interoperability between quantum cloud platforms | Quantum API Gateway [18] is a prototype; no production interoperability standard exists |

| Type of limitation | Description of limitation | Existing solution/gap |
|---|---|---|
| | Significant technical overhead to benchmarking and validation of hardware | Partial solution: cloud platforms (IBM Quantum, Amazon Braket) provide benchmarking tools; cross-platform tooling absent [32] |
| Software | Effort-intensive to write quantum software code for a given cloud service | QFAAS [28] wraps API calls; does not provide genuine co-execution of classical and quantum functions |
| | Unproven software tools for establishing controllable classical–quantum workflows | Partial solution: Weder et al. [40] propose a quantum software lifecycle; toolchain maturity remains low |
| | Accuracy of instruction execution not consistently established | No current solution: Consistency of outputs is an open research problem [24] |
| | Remote calls overhead and queuing latency challenge scalability of QaaS solutions | No current solution: Fundamental to cloud-based QaaS models; latency is architectural [18] [24] |
| | Lack of direct integration between classical and hybrid computing resources | Partial solution: Middleware frameworks [16] and Quantum API Gateway [18] begin to address this; no production-grade solution |
| | C/C++/Python developers lack visibility into information flows and qubit error analysis | No current solution: Educational and tooling gap; domain-specific onboarding materials nascent [32] [11] |

Although classical-quantum variational algorithms [29] have the potential to limit some noise errors inherent to current day quantum computing, the available software tools are unproven when establishing controllable workflows between classical and quantum computers [40]. This means that whether the algorithms employ pre- or post-processing, or employ multiple iterations, the accuracy with which the instructions are executed, and the extent to which the results reflect the intended instructions is not yet consistently established [24].

Beyond the dependence on cloud computing service-orientated architectures and lack of code portability, current hybrid computing solutions also lack direct integration between classical computing software and hybrid computing resources. Domain-specific languages such as Liquid [39], Q# [38], and Qiskit [20] [42] require programmers to develop a deep understanding of quantum logic, including the gate and circuit models.

This poses challenges to software developers well-versed in the high-level language paradigm of C/C++ or Python, leading to the development of classical-quantum hybrid algorithms in either separate stages or through the use of manual workflow interventions instead of end-to-end integration [25], which will likely determine the success of hybrid computing in practice [24].

## 5 FUTURE DIRECTIONS AND DEVELOPMENT AGENDA

Until quantum computing platforms can match the maturity and scale of classical computing platforms, hybrid systems offer a potential long-term solution bridging the capabilities of pure-play classical and quantum computing platforms. However, several developments and advancements in design methodologies, standardised documentation strategies, and resource management capabilities are needed for hybrid systems to be useful in practice. Advances in Artificial Intelligence (AI)-assisted software development are also expected to play an influential role in furthering the capabilities of hybrid systems.

**Methodologies:** For functional, robust hybrid software systems, high-level design methodologies which can map the different computing paradigms of classical and quantum computing effectively are needed [27]. The challenge is fundamental to how design patterns in classical software engineering evolved. These patterns were not invented but were distilled from decades of accumulated practice building and maintaining large systems at scale. Quantum software engineering has no equivalent corpus of production experience to draw from, which means that methodology development cannot simply adapt classical patterns but must proceed alongside the engineering practice it seeks to codify. The specific developments required include:

- Standardised design patterns for hybrid systems
- Domain-specific modelling languages (DSMLs) to help abstract the complexity of integration between classical-quantum systems; and
- Systematic architectural frameworks for different hybrid computing scenarios such as tight vs. loose coupling, and real-time vs batch processing.

**Tools and processes:** Although quantum software engineering is progressing as a discipline, it currently lacks the development procedures, standards, and capabilities available to software engineers conversant with classical computing paradigms. The maturity challenges of quantum software engineering have been articulated in foundational position statements such as the Talavera Manifesto [31]. The challenge is therefore not a shortage of tools but fragmentation: every major vendor and research group has developed its own SDK, testing framework, and simulator, creating a landscape in which skills, workflows, and outputs are largely non-transferable across platforms. A developer proficient in Qiskit cannot straightforwardly apply that expertise to Cirq or to Q#; a test suite written for one backend must be substantially rewritten for another. Until this fragmentation is addressed through open standards and shared abstractions, the cumulative investment in tooling will remain trapped in incompatible silos. Further development of the field requires [11]:

- Standardised phases and practices for hybrid classical-quantum software development cycle

- Requirement engineering approaches adapted or specifically identified to meet the needs of hybrid system development
- Testing and validation frameworks for hybrid system applications; and
- Documentation standards and maintenance strategies so hybrid systems can deliver return on investment in the long-term.

**Resource management:** When the developments in vertical and horizontal hybrid computing are considered, current hybrid systems also lack sophisticated resource management capabilities when their adoption in practice is considered. The difficulty of closing this gap is rooted in the nature of quantum hardware itself: unlike classical processors, quantum hardware cannot be virtualised, time-shared in the conventional sense, or monitored using standard performance metrics. Coherence time constraints impose hard deadlines on circuit execution that have no classical analogue; hardware performance degrades in ways that are invisible to classical monitoring tools; and the cost of QPU access makes naive scheduling strategies economically untenable. Resource management for hybrid systems therefore requires new abstractions that treat quantum coherence time and gate fidelity as first-class scheduling constraints alongside classical metrics such as memory and CPU utilisation. For hybrid systems to be usable in practice, the following advancements are needed [34]:

- Dynamic resource allocation between quantum and classical components
- Performance prediction models for hybrid workloads
- Energy optimisation strategies balancing quantum coherence requirements with classical efficiency; and
- Quality of service (QoS) guarantees in hybrid environments

**AI-assisted software development:** The emergence of large language models (LLMs) also promises to accelerate quantum software development by automating circuit design, improving QUBO formulations, and translating classical algorithms into quantum-compatible code. Since quantum programming requires specialised physics knowledge and quantum programming languages span a wide and fragmented landscape of domain-specific languages [19], LLMs could help reduce barriers to entry by narrowing the expertise gap between software engineers and quantum physicists [27]. Early work on LLM-assisted quantum circuit generation [21] suggests this avenue is active but nascent. The efficiency and effectiveness of LLM-based tools for hybrid computing remains unproven due to limited availability of hardware and the reliance of current development stacks on simulators.

## 6 CONCLUSIONS

This paper examined how the software stack of hybrid classical-quantum computing solutions can be expected to work in practice. Hybrid classical-quantum software development faces a number of challenges i.e. absence of intermediate verification of outputs, lack of interoperability between quantum cloud platforms, unproven software tools for managing workflows between classical and quantum computers, lack of scalability in quantum-as-a-service solutions, and steep learning curve for programming quantum logic through currently available domain-specific languages such as Liquid, Q#, Cirq, and Qiskit.

Although solutions based on classical distributed systems and software engineering paradigms (for example, TOSCA4QC, Quokka, and QFAAS) have the potential to simplify the development of workflows related to the development of hybrid quantum-classical software systems, there are at present limited examples of applications in practice that can demonstrably, consistently exploit the promise of hybrid quantum-classical systems. Most identified limitations remain without a production-grade solution. The solution landscape is nascent and highly fragmented. The near-term priorities for development of functional, robust hybrid software systems therefore include: high-level design methodologies which can map the different computing paradigms of classical and quantum computing; standardised phases and practices for hybrid classical-quantum software development cycle; and sophisticated resource management capabilities aimed at performance prediction, energy optimisation, and quality of service guarantees for vertical and horizontal hybrid computing in practice.

So far hardware limitations of quantum computing have attracted defined, more well-funded research trajectories while software limitations remain structurally unaddressed. This pattern of development is not unique to quantum computing. It is a recurring pattern in the history of the computing field. The transition from batch processing to interactive computing, from centralised to distributed systems, and from procedural to object-oriented programming each involved a period in which hardware capabilities advanced ahead of the software engineering practices needed to exploit them reliably and at scale. In each case, the period of transition was characterised by fragmentation, manual workarounds, and a lack of shared standards, the conditions that prevail for hybrid quantum-classical software today. In each case, the resolution came not from hardware advances alone but from sustained investment in the software engineering foundations: languages, abstractions, tools, standards, and the education of a practitioner community. Hybrid quantum-classical computing is now in that formative period for software.

The implication for the broader computing field is that investment in quantum software engineering, in tooling, interoperability standards, testing methodology, and the education of hybrid software practitioners is as strategically important as investment in quantum hardware, and arguably more tractable in the near term. Progress on the software foundations does not require breakthroughs in physics; it requires the kind of sustained, collaborative engineering effort that the computing community has successfully undertaken before.